\documentstyle[twocolumn,prl,aps,graphicx]{revtex}
\begin{document}
\draft
\title{Limit cycle theory of temporal current self-oscillations 
in sequential tunneling of superlattices}
\author{X. R. Wang$^1$, Z. Z. Sun$^1$, S. Q. Duan$^{1,2}$, 
and Shi-dong Wang$^1$ }
\address{$^1$ Physics Department,
The Hong Kong University of Science and Technology,
Clear Water Bay, Hong Kong, SAR, China}
\address{$^2$ Institute of Applied Physics and Computational 
Mathematics, Beijing, 100088, China}
\date{\today}
\maketitle
\begin{abstract}
A unified theory of the temporal current self-oscillations 
is presented. We establish these oscillations as the 
manifestations of limit cycles, around unstable steady-state 
solutions caused by the negative differential conductance. 
This theory implies that both the generation and the motion 
of an electric-field domain boundary are universal in the sense
that they do not depend on the initial conditions. Under an 
extra weak ac bias with a frequency $\omega_{ac}$, the 
frequency must be either $\omega_{ac}$ or an integer 
fractional of $\omega_{ac}$ if the tunneling current 
oscillates periodically in time, indicating the 
periodic doubling for this non-linear dynamical system.
\end{abstract}

Following the discovery of temporal current self-oscillations 
(TCSOs) in sequential tunneling of superlattices (SLs) 
under a dc bias\cite{grahn2,mimura,kwok}, a large number of 
experimental and theoretical studies have focused on their 
origin and how these oscillations develop from steady-state 
solutions (SSSs). Experimentally, TCSOs have been observed 
in both doped and undoped SLs\cite{grahn2,mimura,kwok}. 
The oscillations can be induced by continuous illumination 
with laser light\cite{kwok} or by a change in doping\cite
{grahn2}. Recently, it has been shown that TCSOs can also 
be induced by applying an external magnetic field parallel 
to SL layers or varying the sample temperature\cite{jwang}. 

Our current theoretical understanding of TCSOs is mainly from 
numerical studies\cite{kastrup}. Early works of Bonilla and his 
co-workers\cite{kastrup} established a correct model for TCSOs. 
Many numerical results and some analyses were also done. 
They simulated and reproduced many experimental results, 
including finding proper model parameters to simulate different 
experimental situations. Great progress in understanding of 
TCSOs had been made because of them. However, a simple physical 
picture did not appear in those early works. The understanding 
at the computational level is the first step, and deep insights 
can be only obtained when the general concepts and principles 
are found. There are also microscopic Green's function 
calculations\cite{wacker}. While the microscopic approach would 
be accurate if all the microscopic parameters and mechanisms 
were known, it remains a challenge to deduce the rules of 
macroscopic behavior from the microscopic details. Recently, a 
clear route to TCSOs developed from SSSs in sequential 
tunneling of SLs was proposed\cite{xrw}: Due to the negative 
differential conductance (NDC)\cite{buttiker}, a SSS is not 
stable. A limit cycle is generated around the unstable SSS 
because of the local repulsion and global attraction in the 
phase space. The system moves along the limit cycle, leading 
to a TCSO. 

Unfortunately, more careful studies\cite{xiong,bonilla} show 
that the simple model used in reference 7 does not have a 
TCSO solution even though it gives the on-set of instability 
conditions of a SSS. Thus, the evidence for limit cycles in the 
TCSO regime is still lacking. In this letter we search for the 
fundamental concepts and general principles for TCSOs, based on 
the correct model of Bonilla and his co-workers\cite{kastrup}. 
{\it We apply the concept of limit cycle, and explain TCSOs as the 
manifestations of limit cycles}. Although numerical results are 
not our emphases in this paper, we show that they can be easily 
understood by using this concept. For example, the generation and 
motion of an EFD boundary do not depend on the initial conditions. 
An EFD-boundary does not necessarily start from the emitter, and 
end up in the collector. Furthermore, limit cycle concept gives us 
a power of prediction. We predict that the generation and motion 
of EFD boundaries are universal, and the frequency of a periodic 
motion under an ac bias must be either the ac bias frequency or 
its integer fractional. The limit cycle is a well-known concept in 
non linear physics. Thus, we find that TCSOs can be understood 
under the general concepts of non linear physics. 
 \begin{figure}
 \vspace{0mm}
  \vbox to 3.5cm {\vss\hbox to 5.0cm
  {\hss\
    {\includegraphics{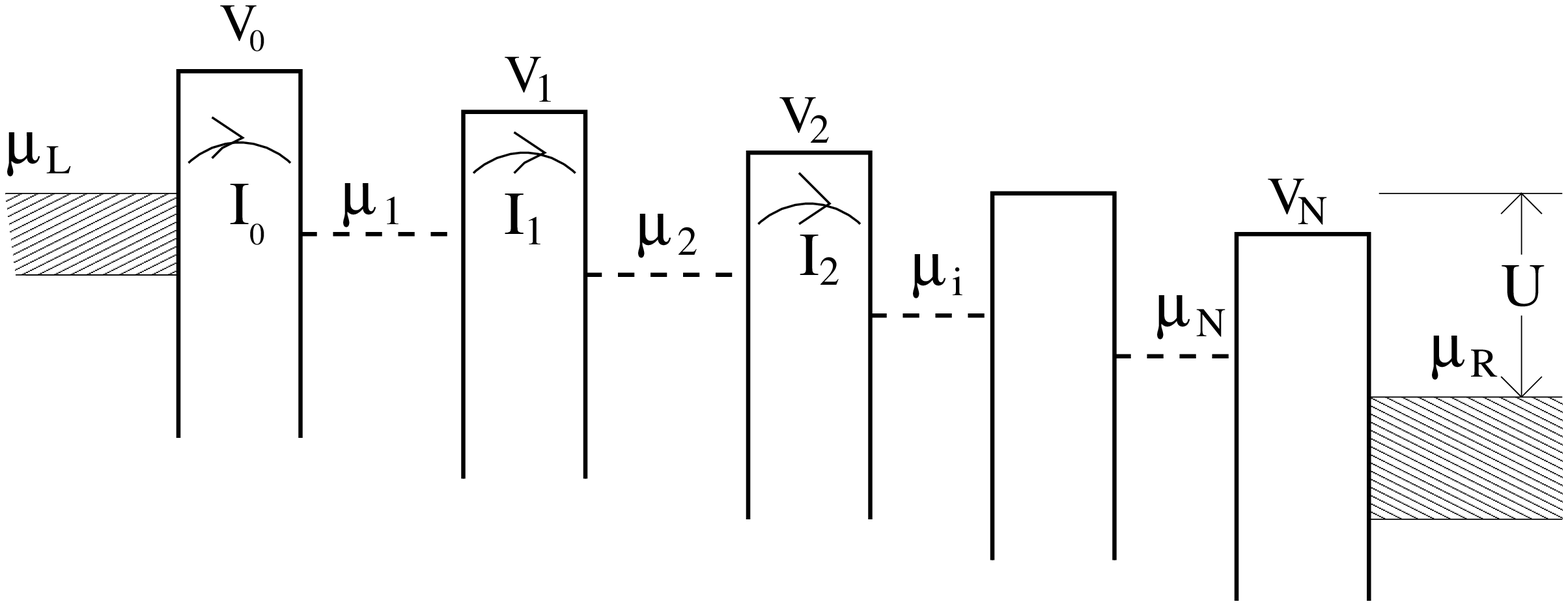}
    }
   \hss}
  }
\vspace{-1mm}
\caption{\label{fig1} Schematic illustration of an SL system. 
$\mu_{i}$ is the local chemical potential of the $i^{th}$ well. 
$\mu_L$ and $\mu_R$ are the chemical potential of left-hand side 
and right-hand side electrodes, respectively. $V_i$ is the bias on 
the $i$th barrier, and $\mu_{L}-\mu_{R}=U$ is the external bias.}
\end{figure}

We consider a system consisting of $N$ quantum wells as shown
schematically in Fig. 1. An external bias $U$ is applied between 
the two end wells. Current flows perpendicular to the SL layers. 
In the sequential tunneling, charge carriers are in local 
equilibrium within each well, so that a chemical potential can be 
defined locally. The chemical potential difference between two 
adjacent wells is called bias $V$ on the barrier between the two 
wells. A current $I_i$ passes through the $i^{th}$ barrier under 
a given bias $V_i$. This current may depend on other parameters, 
such as doping $N_D$. One of the results in reference 7 is that 
a SSS must be unstable if there are two or more barriers being in 
the NDC regime. It is worth pointing out that, although 
this instability result is obtained from an analysis of a 
simplified sequential tunneling model, it is generally true. 
Without losing generality, we assume that only barriers 1 and 2, 
which create well 2 shown in Fig. 1, are under a NDC regime for 
a SSS. Assume the chemical potential in well 2 increases a little 
bit due to a fluctuation. Then bias $V_1$ on barrier 1 decreases 
while $V_2$ increases. Because both of the barriers are under 
NDC regime, charge carriers flow more into well 2 through 
barrier 1 while less carriers flow out of it, leading 
to a further increase of the chemical potential in well 2. 
This drives the system away from the SSS, i.e. instability. 

Following reference 5, the dynamics of the system is governed by 
the discrete Poisson equations 
\begin{equation}
\label{poisson}
k(V_i-V_{i-1})=n_i-N_{D}, \qquad i=1,2,\ldots N,
\end{equation}
and the current continuity equations 
\begin{equation}
\label{charge}
J=k\frac{\partial V_{i}} {\partial t} + I_i
, \qquad i=0,1,2,\ldots N 
\end{equation}
where $k$ depends on the SL structure and its dielectric constant. 
$n_i$ is the electric charge in the $i^{th}$ well. 
In Eq. (\ref{poisson}), a same doping in all wells is assumed. 
$I_i$ is, in general, a function of $V_i$ and $n_i$. It can be shown
\cite{xiong} that all SSSs are stable if $I_i$ is a function of $V_i$ 
only. On the other hand, a SSS may be unstable\cite{kastrup} if one 
chooses $I_i=n_{i}v(V_{i}),$ where $v$ is a phenomenological drifting 
velocity which is, for simplicity, assumed to be a function of $V_i$ 
only. The constraint equation for $V_i$ is
\begin{equation}
\label{bias}
\sum_{i=0}^{N}V_i=U.
\end{equation}
Previous studies\cite{kastrup} proved that this model is capable of 
describing TCSOs. To close the equations, a proper boundary condition 
is needed. It is proper to assume a constant $n_0$, $n_0=\delta N_{D}$
with $\delta$ as a model parameter, 
if the carrier density in the emitter is much larger than those in 
wells, and its change due to a tiny tunneling current is negligible. 
However, it is mathematically equivalent to other boundary conditions
used in literature\cite{kastrup}. Our goal is to show that the limit 
cycle is one of the most important features in this widely studied model.

According to our theory, NDC is essential for TCSOs. Thus, a TCSO 
can only occur when there is a negative differential velocity in 
$v(V)$\cite{kastrup}. There are many ways of choosing it. One can 
assume $v$ being a sum of a series of Lorentzian functions if NDC 
is due to the resonance tunneling between discrete quasi-localized 
states of wells. One may also choose a piecewise linear function 
in order to make an analytic investigation easy. We shall assume 
$v(V)$ as the sum of two Lorentzian functions, 
$v(V)=0.0081/[(V/E-1)^{2}+0.01]+0.36/[(V/E-2.35)^{2}+0.18]$. 
This $v$ has two peaks at $V=E$ and $V=2.35E$. A negative 
differential velocity exists between $V=E$ and about $V=1.3E$. 
Thus, $E$ can be used as a natural unit of bias, and $1/v(E)$ as 
that of the time (the lattice constant is set to be 1). For 
$N=40$, $U=43.6E$, $N_{D}=0.095kE$, and $\delta=1.001$, the set 
of equations has TCSO solutions. We solved the set of equations 
by using the Runge-Kutta method. Initially, the external bias is 
randomly distributed. Quickly, $V_{i}$ reaches a stable state 
{\it which does not depend on the intial conditions}. 
Figure 2 is the projection of the stable bias-trajectory in 
$V_{5}-V_{38}$ phase plane. $V_5$ is in the low EFD while $V_{38}$ 
is in the high EFD here. Clearly, one obtains a closed {\it 
isolated} curve indicating a limit cycle. The current oscillation 
period, the inverse of system intrinsic frequency $\omega_0$, is 
the time that the system needs to move around the cycle once. 
The inset is the phase diagram of the TCSOs in $U$-$N_D$ plane with 
all other parameters unchanged, where $N_D$ is in units of $kE$. 
The shadowed area corresponds to the TCSO regime. Of course, the 
diagram depends on the values of $\delta$, $N$, and function of 
$v$.
 \begin{figure} 
 \begin{center}
 \includegraphics[width=7.0cm, height=4.5cm]{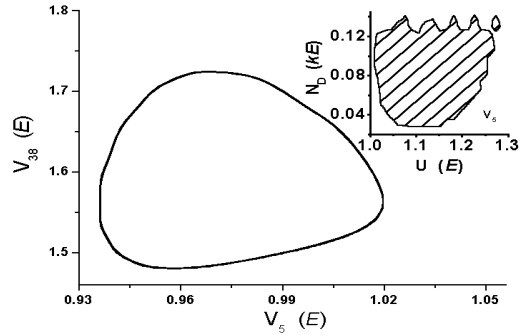}
 \end{center} 
\caption{\label{fig2} The trajectory of the system in phase plane 
$V_5-V_{38}$ in a TCSO regime. The closed curve is the projection 
of the limit cycle in the plane. The bias is in units of $E$. 
The inset is the phase diagram in $U$-$N_D$ plane with all other 
parameters unchanged, where $N_D$ is in units of $kE$.
The system inside the shadowed area will have a TCSO solution while 
it has a static current-voltage characteristic outside this area. }
\end{figure}

Although it is known that TCSOs are accompanied by the motion of EFD 
boundaries, how EFD boundaries are generated and propagate inside 
SLs were debated\cite{kastrup}. According to the limit cycle picture, 
the charge accumulation (depletion) is responsible to the creation of 
an EFD boundary. Charge carriers are accumulated (depleted) in a 
particular well because of an imbalance of carriers flowing in and 
flowing out. This imbalance is caused by NDC as we argued early. 
Thus an EFD boundary can start at any well and oscillates inside a SL. 
Furthermore, {\it the limit cycle picture means that EFD boundaries 
should not depend on the initial conditions, but are completely 
determined by the limit cycles around unstable SSSs}.

To demonstrate the correctness of our picture, we locate numerically 
the position of the EFD boundary in the calculation that gives Fig. 
2. Figure 3 is the evolution of the boundary. It reaches a 
stable oscillating state quickly. One can see that the EFD boundary 
oscillating between wells 26 and 37 in the SL of total 40 wells. 
The inset is the field distribution across the SL at points a, b, 
c, and d in Fig. 3. The bias $V_5$ ($V_{38}$) of the low (high) EFD 
moves up and down as the EFD boundary oscillates inside the SL. 
The stable oscillation state, a manifestation of a limit cycle, 
does not change when different initial conditions are used. 
In this sense both generation and motion of an EFD boundary 
are universal. Although early numerical calculations\cite{kastrup} 
might have already implied that an EFD boundary can start in an 
interior well and oscillate inside a SL. It may not be easy for 
theories like that of reference 5 to explain this universal property. 
 \begin{figure}[hb]
 \begin{center}
 \includegraphics[width=7.cm, height=4.5cm]{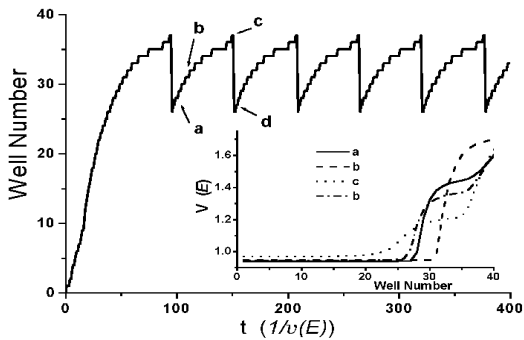}
 \end{center} 
\caption{\label{fig3} The time evolution of an EFD boundary. The 
parameters are the same as those for Fig. 2. The boundary is 
oscillating between well 26 and well 37. The stable oscillation 
does not depend on the initial condictions. The inset is the 
electric field distribution at points a, b, c, and d in Fig. 3. 
}
\end{figure}

Except a few of attempts\cite{bulashenko,scholl} which were 
mainly on the numerical aspects dealing with non-periodic 
time-dependent tunneling current, most theoretical studies 
have not considered TCSOs under the influence of an ac bias 
due to the lack of a clear physical picture such that one 
would not be able to analyze its effects. The limit 
cycle theory offers a way of analyzing the ac bias effect. 
Applying a small extra ac bias, it affects TCSOs through 
perturbing the system trajectories. In the case that the 
limit cycles still exist and are stable, or in other words, 
the tunneling current can oscillate periodically with 
frequency $\omega$, the ac bias on the system should return 
to its starting value after the system completes one-round 
motion on the limit cycle. It means $\omega_{ac}/\omega=n=
integer$. A weak ac bias cannot greatly change the 
evolution trajectory of the system in the phase space. 
The consequences are as follows: 
a) Current oscillation frequency $\omega$ cannot be much 
greater than the intrinsic frequency. Thus, at high 
ac bias frequency ($\omega_{ac}>>\omega_{0}$), 
it is impossible for the bias to deform the limit cycle 
slightly such that the time for the system moving around 
the cycle once to be the same as the period of the ac bias. 
b) In a case that the system cannot deform itself to match 
$\omega_{ac}$, a trajectory may become a closed curve 
after several turns in the phase space. 
Therefore, $\omega_{0}/\omega=m=integer$. 
Indeed, figure 4 is the limit cycles in 
phase plane $E_5$-$E_{38}$ under an extra ac bias $V_{ac}
\cos(\omega_{ac}t)$ with $V_{ac}=0.44E$ and 
$\omega_{ac}=2\omega_0$ (dot line), $3\omega_0$ (solid line) 
while the rest of parameters remain the same as those for 
Figs. 2-3. The right (left) inset is the current-time curve 
for $\omega_{ac}=2\omega_0$ ($3\omega_0$) after the current 
oscillation becomes stable. The Fourier transformation 
shows the current oscillation frequency $\omega$ being 
$\omega_0$. This is the solution of $\omega_{ac}/\omega=n=integer$
and $\omega_{0}/\omega=m=integer$ for the smallest possible 
$n$ (=2, 3) and $m $ (=1). 
 \begin{figure}[hb]
 \begin{center}
 \includegraphics[width=7.cm, height=4.5cm]{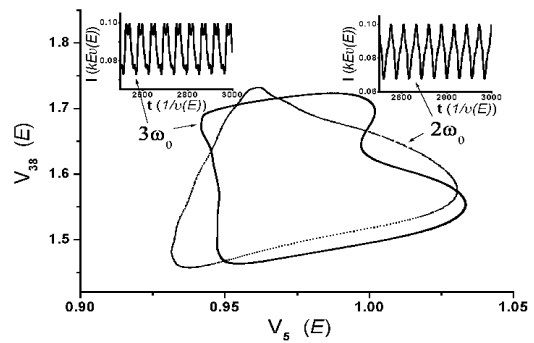}
 \end{center}
\caption{\label{fig4} The limit cycles in phase plane $V_5$-$V_{38}$
under an extra weak ac bias $V_{ac}\cos (\omega_{ac}t)$ with 
$\omega_{ac}$ being $2\omega_0$ and $3\omega_0$. The inset on the 
right (left) is the current-time curve for $\omega_{ac}=2\omega_0$
(=$3\omega_0$). $V_{ac}=0.44E$, and all other parameters are the 
same as those for Figs. 2 and 3. The time is in units of $1/v(E)$. 
biases are in units of $E$. The current frequencies are 
$\omega_{ac}/2=\omega_0$ and $\omega_{ac}/3=\omega_0$, respectively.
}
\end{figure}

For $\omega_{ac}/\omega_0$ being a non-integer rational number, 
$\omega$ should be different from $\omega_0$ according to the 
rules of $\omega_{ac}/\omega=n=integer$ and $\omega_{0}/\omega=
m=integer$. For example, $\omega_{ac}/\omega_0 = 1/q$ with 
$q=$integer, then the limit cycle can deform itself in such a way 
that it becomes a closed curve after $q$ turns in the phase space, 
corresponding to $n=1$ and $m=q$. In this case, the current 
oscillates with $\omega_{ac}$. The solid line in figure 5 is the 
numerical results of the limit cycle in phase plane $V_5$-$V_{38}$ 
for $\omega_{ac}=\omega_0/3$ while the rest of parameters are kept 
the same as those for Figs. 2 to 4. Indeed, the limit cycle, which 
does not depend on the initial conditions, makes $q=3$ turns in the 
phase plane as expected. The left inset is the corresponding 
tunneling current evolution curve. The Fourier transformation of 
the current evolution indeed shows $\omega=\omega_{ac}$. 
For $\omega_{ac}/\omega_0 =2.5$, our rules predict the $\omega
=\omega_0/2=\omega_{ac}/5$, corresponding to $n=5$ and $m=2$. 
This result is verified by the numerical calculation as displayed 
in figure 5 (dot line). As before, all other parameters are kept 
the same as those for Figs. 2 to 4. The right inset is the 
corresponding tunneling current evolution curve. The limit cycle 
is a two-turn closed curve in phase plane $V_5$-$V_{38}$. The 
Fourier transformation of the current evolution demonstrates that 
the current oscillates with frequency $\omega_0/2=\omega_{ac}/5$.
We would like to make following remarks.
a) $\omega=\omega_{ac}$ for $\omega_{ac}=\omega_0/q$ cannot be true 
for all $q$ because $\omega$ should approach $\omega_0$ in the limit 
$\omega_{ac}\rightarrow 0$. 
b) We have considered only periodic responses of the tunneling 
current. {\it It does not rule out other more complicated behaviors}. 
In fact, there are reports\cite{bulashenko,scholl,zhang} of chaotic 
current-time behavior under a combined dc and ac biases. 
c) Vary external parameters, the size and shape of a limit cycle 
should change in general. Thus $\omega_0$ can shift. As a consequence,
$\omega_{ac}/\omega=n$ may be quite robust against $\omega_{ac}$.
When $\omega_{ac}$ is very close to $\omega_0$, it may be possible 
for the limit cycle to deform itself in such a way that $\omega_0$ 
shifts to $\omega_{ac}$. In this case, the current shall oscillate 
with frequency $\omega_{ac}$. Our preliminary results indeed show so. 
How much $\omega_0$ can shift depends on the magnitude of $V_{ac}$ 
and $\omega_{ac}$. The detail analysis in this regime will be our 
future direction in this project.
 \begin{figure}[hb]
 \begin{center}
 \includegraphics[width=7.cm, height=4.5cm]{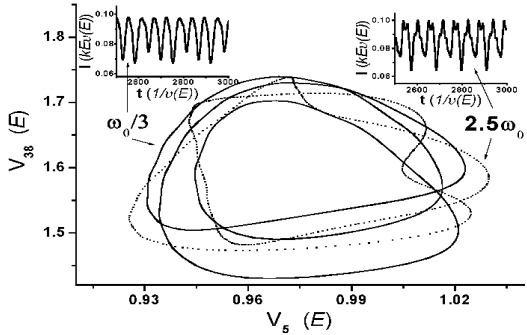}
 \end{center}
 \caption{\label{fig5} The limit cycles in phase plane $V_{5}-V_{38}$ 
under an extra weak ac bias $V_{ac}\cos (\omega_{ac}t)$ with 
$\omega_{ac}$ being $\omega_0/3$ (solid line) and $2.5\omega_0$
(dot line). The inset on the left (right) is the curve of current 
vs. time for $\omega_{ac}=\omega_0/3$ (=$2.5\omega_0$). 
All other parameters are the same as those for Fig. 4.
The current frequency equals to $\omega_{ac}$ ($\omega_{ac}/5$) for
$\omega_{ac}=\omega_0/3$ (=$2.5\omega_0$).
}
\end{figure}

In summary, we show that the trajectories of SLs in their phase 
spaces are closed curves when TCSOs occur. These closed curves 
do not depend on the initial conditions. In other words, the closed 
trajectories are isolated. Thus we show numerically the existence of 
the limit cycles, and TCSOs can be understood as the manifestations 
of limit cycles. Just like many other non-linear dynamical systems, 
TCSOs are governed by the properties of the unstable SSSs. 
According to this theory, the generation and motion of an EFD 
boundary are also the properties of unstable SSSs. An EFD boundary 
does not necessarily need to start from the emitter. It can start 
from an interior well of a SL, and it then oscillates inside the SL. 
They are universal in the sense that they do not depend on the 
initial conditions. This universal property may not be so obvious 
in previous theories\cite{kastrup}. We have also investigated the 
effects of a small extra external ac bias on TCSOs. 
We find that the tunneling current will oscillate periodically when 
the ac bias frequency $\omega_{ac}$ is commensurate with the system 
intrinsic frequency $\omega_0$. The current frequency 
equals either $\omega_{ac}$ or $\omega_{ac}/n$, where $n$ 
is an integer, showing periodic doubling which is a general 
phenomenon in non linear dynamical systems. 

X.R.W. would like to thank Prof. J. N. Wang and Prof. W. K. Ge for 
many comments. This work is supported by UGC, Hong Kong.

\end{document}